\documentclass{aa}  
\usepackage{graphicx}
\usepackage{txfonts}
\begin{document}
   \title{VLT Spectropolarimetry of the Type Ia
   SN~2005ke}
  \subtitle{A step towards understanding subluminous events\thanks{Based on 
   observations made with ESO Telescopes at the
   Paranal Observatory under Program IDs 076.D-0177 and 076.D-0178.}}
   \author{F. Patat\inst{1}
   \and
     P.~H\"oflich\inst{2}
   \and
    D. Baade\inst{1}
    \and
    J.~R. Maund\inst{3}
    \and
    L. Wang\inst{4}
    \and
    J.~C. Wheeler\inst{5}
}

   \offprints{F. Patat}

   \institute{European Organisation for Astronomical Research in the 
	Southern Hemisphere (ESO), Karl-Schwarzschild-Str. 2,
              85748, Garching b. M\"unchen, Germany\\
              \email{fpatat@eso.org}
              \and
	      Department of Physics, Florida State University, Tallahassee,
	      Florida 32306-4350, USA             
              \and
              Queen's University Belfast, Royal Society Research Fellow,
              Belfast, BT7 1NN, UK
              \and
              Department of Physics, Texas A\&M University, College Station, 
	      Texas 77843, USA
              \and
              Department of Astronomy and McDonald Observatory, 
              The University of Texas at Austin, Austin, TX 78712, USA
             }

   \date{Received March, 2012; accepted May 2012}

\abstract
{}
{In this study we try to answer the question whether or not
  subluminous Type Ia supernovae have additional distinctive
  properties when examined from the point of view of the explosion 
  geometry. 
}
{We have performed optical spectropolarimetric observations of the
  Type Ia SN~2005ke at 3 epochs (days $-$8, $-$7, and +76).  The
  explosion properties are derived by comparing the data to explosion and
  radiation transfer models.
}
{The SN shows polarimetric properties that are very similar to the
  only other subluminous event for which spectropolarimetry is
  available, i.e.  SN~1999by. The data present a very marked dominant
  axis, which is shared by both the continuum and lines such as
  \ion{Si}{ii} $\lambda$6355, suggesting that the relatively large,
  global asymmetry is common to the photosphere and the line-forming
  region. The maximum polarization degree observed in the \ion{Si}{ii}
  $\lambda$6355 absorption reaches 0.39$\pm$0.08\%. At variance with
  what is seen in core-normal Type Ia, SN~2005ke displays
  significant continuum polarization, which grows from the blue to the
  red and peaks at about 7000 \AA, reaching $\sim$0.7\%. The
  properties of the polarization and flux spectra can be understood
  within the framework of a subluminous delayed-detonation (DD), or
  pulsating DD scenario, or WD mergers. The difference in appearance with respect to core-normal SNe~Ia is
  caused by low photospheric temperatures in combination with layers
  of unburned C, and more massive layers of the products of explosive C and O
  burning. The comparatively high level of continuum polarization is
  explained in terms of a significant global asymmetry ($\sim$15\%),
  which is well reproduced by an oblate ellipsoidal geometry within
  the general context of a delayed-detonation explosion. 
 }
{Our results suggest that SN~2005ke arose either from a
  single-degenerate system in which the WD is especially rapidly rotating,
  close to the break-up velocity, or from a double-degenerate merger.
  Based on the current polarization data, we cannot distinguish
  between these two possibilities. Possible tests are discussed.}

\authorrunning{F. Patat et al.}
\titlerunning{VLT spectropolarimetry of the Type Ia SN~2005ke}

   \maketitle
%

\section{\label{sec:intro}Introduction}

Subluminous Type Ia Supernovae (SNe Ia) are still a puzzle in our
understanding of this important class of explosive events (Hillebrandt
\& Niemeyer \cite{hn00}). Although they contribute to ``only'' 15\% of all
Type Ia events (Li et al. \cite{li11}), understanding their progenitors
and the mechanism that powers their explosions is fundamental to fully
grasp the Type Ia supernova phenomenon (see, for instance, Pakmor et
al. \cite{pakmor10}).

These objects, commonly dubbed ``91bg-like'' after the prototypical
event SN~1991bg (Filippenko et al. \cite{flipper92}, Leibundgut et
al. \cite{leib93}), deviate quite significantly from the behavior
defined by core-normal events (Branch et
al. \cite{branch06}). Photometrically, they are about 2 magnitudes
fainter, display rapidly declining light curves with narrower peaks,
lack a secondary maximum in the NIR bands, have redder colors at
maximum light, and obey a different relation between maximum
luminosity and light curve shape (Garnavich et
al. \cite{garnavich04}). Spectroscopically, they show cool spectra,
characterized by intermediate-mass elements, particularly strong
\ion{O}{i} and \ion{Ti}{ii} absorptions, and lower expansion
velocities. In addition, both iron-group elements and silicon are
spread over velocities spanning a large fraction of the ejecta,
implying very substantial mixing (Taubenberger et
al. \cite{tauben08}), as opposed to what happens in core-normal
objects. The estimated mass of $^{56}$Ni is as low as 0.1 M$_\odot$
(Stritzinger et al. \cite{stritzinger06}).

All these facts pose a challenge and call for a different explosion
mechanism for subluminous SNe Ia than for core-normal SNe Ia.
As of today, although several theoretical scenarios have
been put forward (see Pakmor et al. \cite{pakmor11} for a recent
review), we do not have a clear idea about the origin of this
sub-class of Type Ia SNe. On the observational side, there are
indications that these objects form a rather homogeneous and distinct
class; however, there is growing evidence that transition objects,
sharing properties of both subluminous and core-normal SNe, d
exist (Maguire et al. \cite{maguire}).

\begin{figure}
\centering \includegraphics[width=8cm]{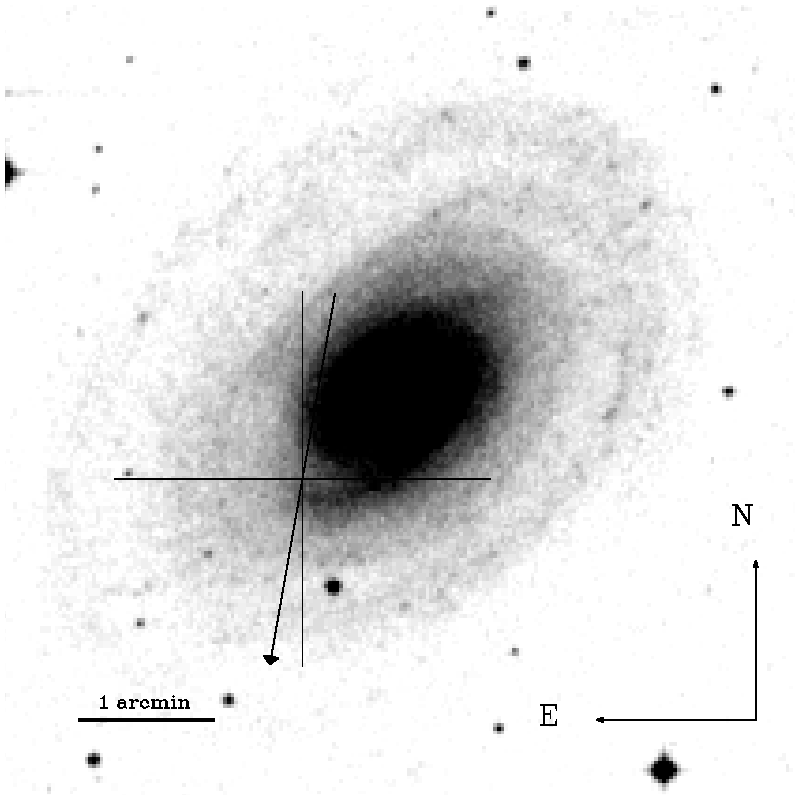}
\caption{\label{fig:fchart}A DSS image of NGC~1371. The location of 
SN~2005ke is marked by the cross. The arrow indicates the ISP
position angle ($\theta_{ISP}$=170 degrees; see Sec.~\ref{sec:specpol}).}
\end{figure}

An additional channel for investigating the diversity of SNe is
polarimetry, which enables the study of the explosion geometry (Wang
\& Wheeler \cite{WW08}). Well studied 91bg-like events are quite rare,
both because of their lower luminosity and their intrinsic lower
rate. For the same reasons, subluminous events studied through
spectropolarimetry are exceptionally rare (Wang \& Wheeler
\cite{WW08}). More precisely, so far the geometrical properties of the
explosion have only been studied for one such SN, i.e. SN1999by, which
showed very distinct features (Howell et al. \cite{howell01}).

In this paper we present a spectropolarimetric study of another
91bg-like event, SN~2005ke. This SN was discovered by Baek, Prasad \&
Li (\cite{baek}) on 13 November 2005 (UT) in the Virgo Cluster spiral
galaxy NGC~1371 (see Fig.~\ref{fig:fchart}), located at a distance of
25.8$\pm$3.2 Mpc (Wood-Vasey et al. \cite{woodvasey}) and receding
with a velocity v$_{gal}$=1463 km s$^{-1}$ (Koribalski et
al. \cite{koribalski}). A few days later, the object was classified as
an underluminous Type Ia event caught before maximum light (Patat et
al. \cite{class}). Spectroscopic and photometric data were presented
by Immler et al. (\cite{immler}), Bufano et al. (\cite{bufano}),
Folatelli et al. (\cite{folatelli}), and Contreras et
al. (\cite{contreras}), to which we refer the reader for the general
properties of this object.

\begin{figure}
\centering
\includegraphics[width=9cm]{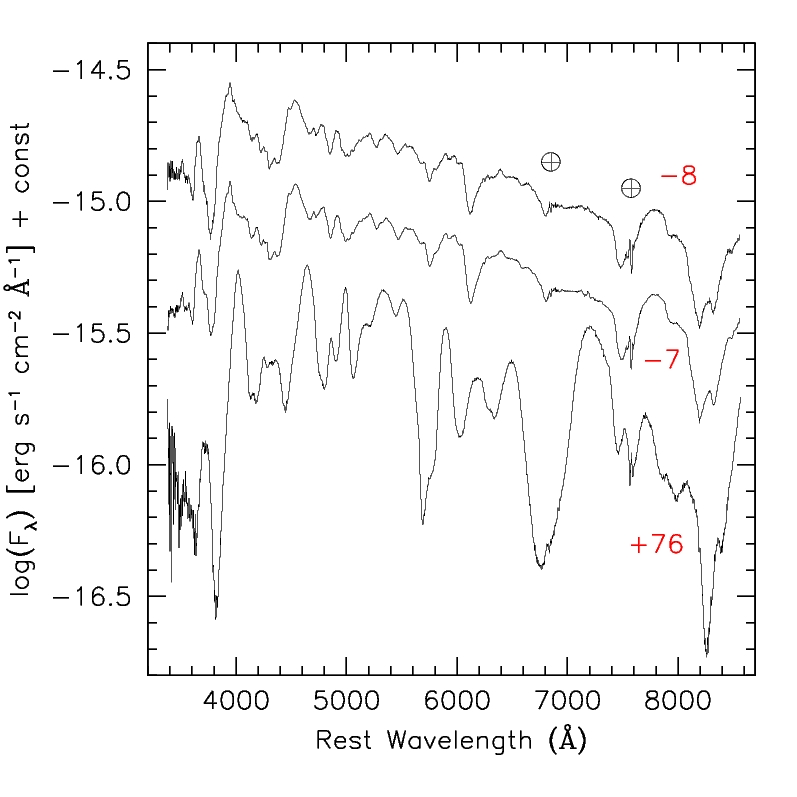}
\caption{\label{fig:evol}Spectroscopic evolution of SN~2005ke during the
epochs covered by the observations presented in this paper. Phases
refer to $B$ maximum light (JD=2,453,699.16; Folatelli et
al. \cite{folatelli}). For presentation the spectrum of the second
epoch is shifted by $\log$F=$-$0.5.}
\end{figure}

The paper is organized as follows. In Sect.~\ref{sec:obs} we present
and discuss our observations and data reduction. The spectroscopic
evolution during the phases covered by our observations is analyzed in
Sect.~\ref{sec:evol}, while Sect.~\ref{sec:specpol} presents the
spectropolarimetric data. The data are compared to explosion and
radiation transfer models in Sect.~\ref{sec:models}. In
Sect.~\ref{sec:disc} we discuss our results, while in
Sect.~\ref{sec:concl} we summarize the conclusions.

\section{\label{sec:obs}Observations and Data Reduction}

\begin{table}
\centering
\tabcolsep 0.8mm
\caption{\label{tab:obs}Log of VLT-FORS1 spectropolarimetric observations 
of SN~2005ke.}
\begin{tabular}{cccccc}
\hline
Date  & MJD                & phase       & $V$ & airmass     & exp. time\\
(UT)  & (JD-2,400,000.5)   & (days)      & (*) &  (average)  & (seconds) \\
\hline
2005-11-16 & 53690.10 & $-$8.6  & 15.2 &1.1  & 4$\times$300\\
2005-11-17 & 53691.10 & $-$7.6  & 15.0 &1.2  & 4$\times$(300+600)\\
2006-02-08 & 53775.06 &  +76  & 17.7 &1.2   & 4$\times$900\\
\hline
\multicolumn{6}{l}{(*) $V$ magnitudes are derived from Contreras et al. 
(\cite{contreras}).}
\end{tabular}
\end{table}

\begin{figure}
\centering
\includegraphics[width=9cm]{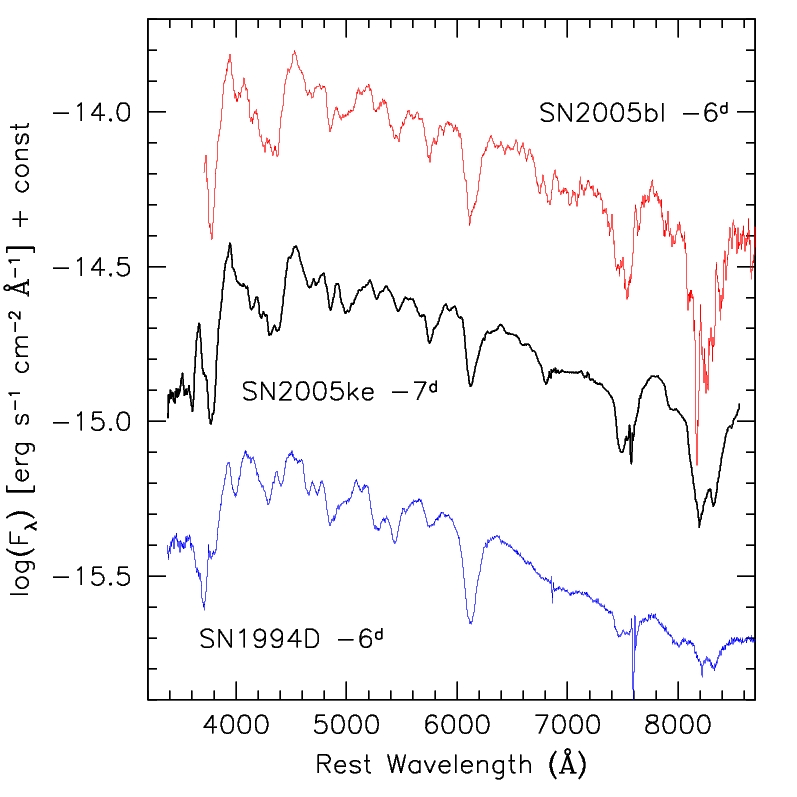}
\caption{\label{fig:comp}Spectroscopic comparison of subluminous SNe
  2005ke, 2005bl (Taubenberger et al. \cite{tauben08}) and
  core-normal SN~1994D (Patat et al. \cite{patat06}) at $-$7/$-$6 days.}
\end{figure}

We have observed SN~2005ke on 3 different epochs, using the FOcal
Reducer/low-dispersion Spectrograph (hereafter FORS1), mounted at the
Cassegrain focus of the ESO--Kueyen 8.2m telescope (Appenzeller et al.
\cite{appenzeller}). In this multi-mode instrument, equipped with a
2048$\times$2048 pixel (px) TK2048EB4-1 backside-thinned CCD,
polarimetry is performed introducing a Wollaston prism
(19$^{\prime\prime}$ throw) and a super-achromatic half-wave plate. In
order to reduce some known instrumental problems (see Patat \&
Romaniello \cite{patat06}) we have always used 4 half-wave plate (HWP)
angles (0, 22.5, 45 and 67.5 degrees). The SN was observed at two
pre-maximum phases ($-$8.6 and $-$7.6 days) and one additional epoch (two
and a half months past maximum light). Exposure times ranged from 5 to
15 minutes per plate angle and, on the second epoch, the sequence was
repeated in order to increase the signal-to-noise ratio. All spectra
were obtained with the low-resolution G300V grism coupled to a 1.1
arcsec slit, giving a spectral range 3300-8600 \AA, a dispersion of
$\sim$2.9 \AA\/ pixel$^{-1}$ and a resolution of 12.4 \AA\/ (FWHM) at
5800 \AA. Data were bias, flat-field corrected and wavelength
calibrated by means of standard tasks within IRAF\footnote{IRAF is
  distributed by the National Optical Astronomy Observatories, which
  are operated by the Association of Universities for Research in
  Astronomy, under contract with the National Science
  Foundation.}. The RMS error on the wavelength calibration is about
0.7 \AA. The wavelength scale was corrected to the rest-frame using
the host galaxy recession velocity (1463 km s$^{-1}$; Koribalski et
al.  \cite{koribalski}).

The ordinary and extraordinary beams were processed separately. Stokes
parameters were computed by means of specific routines written by us,
and error estimates were performed following the prescriptions
described by Patat \& Romaniello (\cite{patat06}), while the HWP
zeropoint angle chromatism was corrected using tabulated data (Jehin,
O'Brien \& Szeifert \cite{fors}). In order to increase the
signal-to-noise ratio, multiple data sets obtained at the same epoch
were combined, and the final Stokes parameters binned in $\sim$52
\AA\/ wide bins (20 pixels).

Flux calibration was achieved through the observation of
spectrophotometric standard stars with the full polarimetric optics
inserted (HWP angle set to 0 degrees). Instrumental polarization and
the position angle offset were checked by observing polarized and
unpolarized standard stars, obtained within the FORS1 calibration
plan.

The log of observations is reported in Table~\ref{tab:obs}, where the
SN phases were computed with respect to the $B$ maximum light
(JD=2,453,699.16; Folatelli et al. \cite{folatelli}).

\begin{figure}
\centering
\includegraphics[width=9cm]{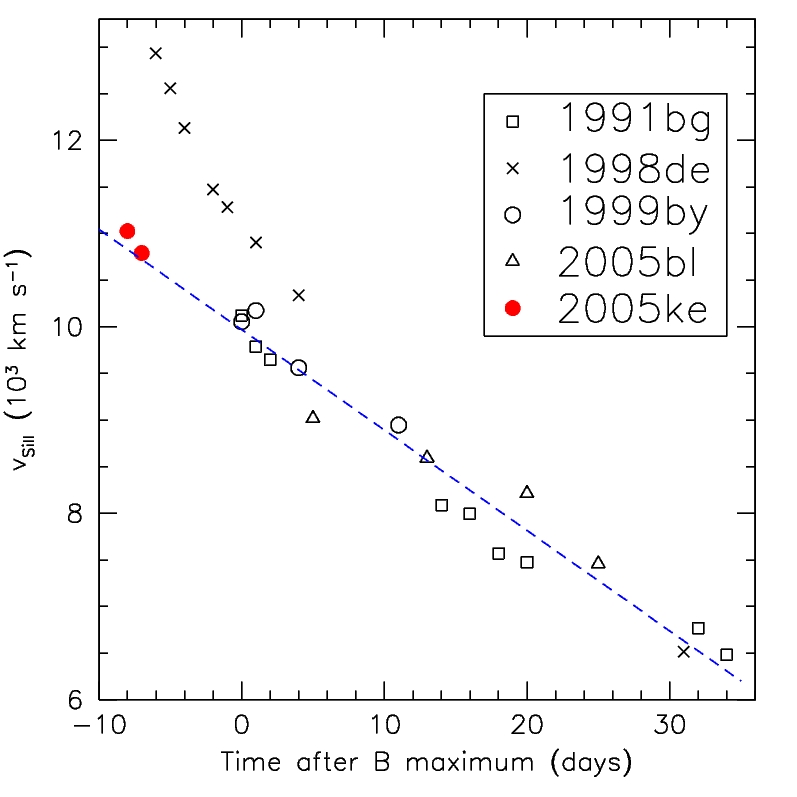}
\caption{\label{fig:vel} Evolution of \ion{Si}{ii} $\lambda$6355 for a
  compilation of underluminous Type Ia: 1991bg (Turatto et
  al. \cite{turatto}), 1998de (Modjaz et al. \cite{mariam}; Matheson
  et al. \cite{matheson08}), 1999by (Vinko et al. \cite{vinko}; Howell
  et al. \cite{howell01}), and 2005bl (Taubenberger et
  al. \cite{tauben08}). The dashed line is a best fit to the data
  after excluding SN1998de.}
\end{figure}

\section{\label{sec:evol}General properties of SN~2005ke}

From the classification spectrum it was clear that SN~2005ke was a
subluminous Type Ia event (Patat et al. \cite{patat05}), very similar
to SN~1999by (Howell et al. \cite{howell01}; Garnavich et
al. \cite{garnavich04}). The characteristic \ion{Ti}{ii} feature at
about 4200 \AA\/ was clearly detected in the pre-maximum spectra.  The
spectroscopic evolution of SN~2005ke during the epochs discussed in
this paper is presented in Fig.~\ref{fig:evol}.  The close resemblance
to another well-studied subluminous SN Ia 2005bl (Taubenberger et
al. \cite{tauben08}) is illustrated in Fig.~\ref{fig:comp}, where for
comparison we have also plotted the spectrum of the core-normal
SN~1994D at the same pre-maximum epoch. The expansion velocity,
deduced from the minimum of the \ion{Si}{ii} $\lambda$6355 absorption
trough on day $-$7, is 11,300 km s$^{-1}$, very similar to the values
measured for the subluminous SN~2005bl (Taubenberger et
al. \cite{tauben08}). On the following day, the velocity dropped to
about 10,900 km s$^{-1}$. These values are compared to those published
for other subluminous Ia events in Fig.~\ref{fig:vel}. Although 
only early measurements are available for
SN~2005ke, and therefore no firm
conclusion can be drawn, this object seems to conform to the behavior
shown by the other 91bg-like events. A best fit to the combined data
of SN~1991bg, 1999by and 2005ke gives a velocity gradient
$\dot{\mbox{v}}$=106$\pm$9 km s$^{-1}$ day$^{-1}$. 

\begin{figure*}
\centering
\includegraphics[width=12cm]{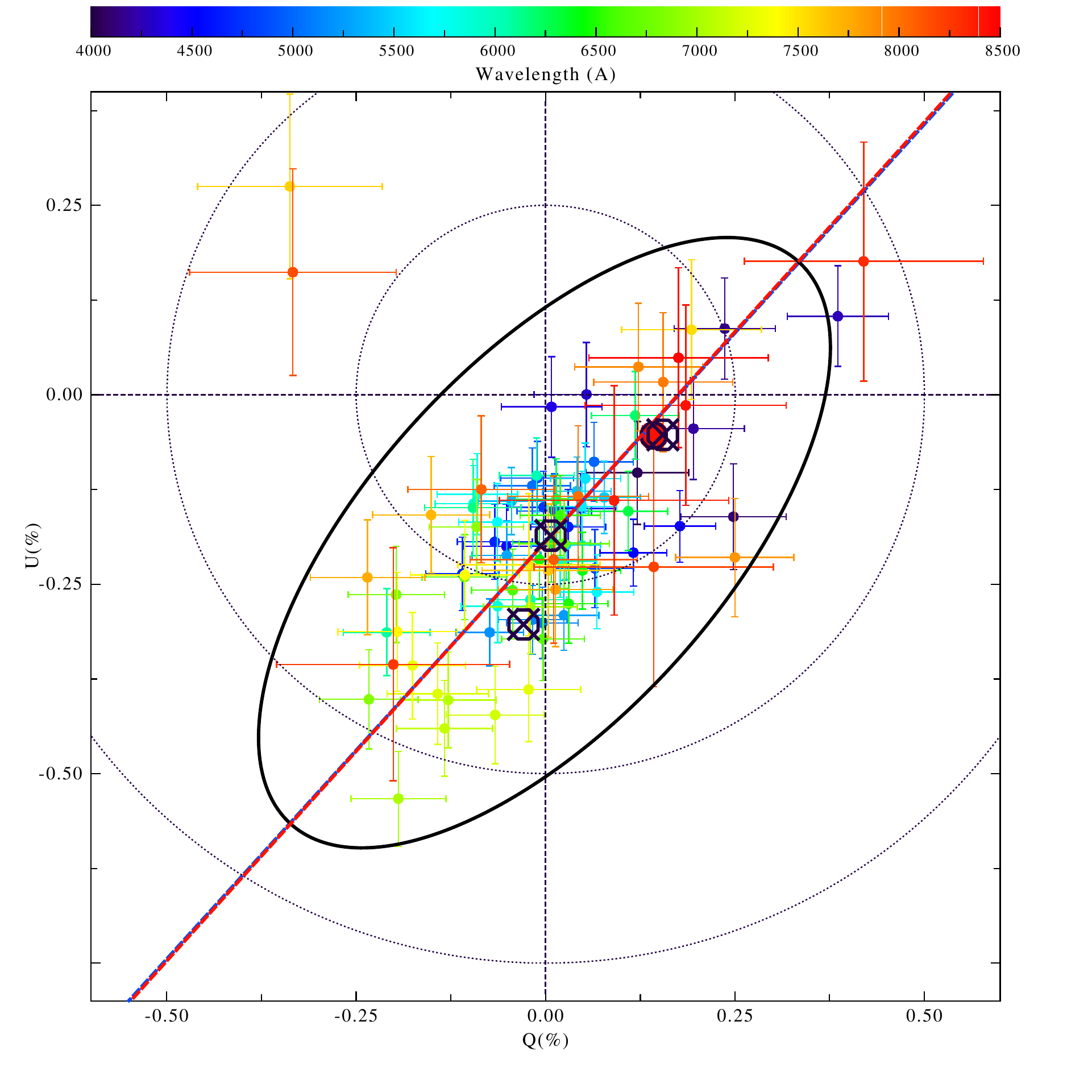}
\caption{\label{fig:dominant} Combined spectropolarimetry of SN~2005ke
  (November 16 and 17, 2005).  The large filled circle marks the ISP
  estimated from the late-time data (see Sect.~\ref{sec:isp}), while
  the circled crosses indicate B, A and C ISP estimates (from right to
  left respectively; see Table~\ref{tab:isp}).  The dashed line traces
  the best fit dominant axis, while the ellipse indicates the axial
  ratio and orientation of the data (see Sect.~\ref{sec:pol}). For
  presentation, the size of the ellipse was arbitrarily scaled. The
  dotted circles indicate polarization levels of 0.25, 0.50 and
  0.75\%.}
\end{figure*}

SN~1998de displays higher
velocities in the pre-maximum epochs, while a month past maximum light
it falls very close to the relation derived from the previous objects.
Its average velocity gradient is $\sim$175 km s$^{-1}$
day$^{-1}$. Although this indicates a possible break in the
homogeneity otherwise seen in this sub-class of SNe Ia (Taubenberger
et al. \cite{tauben08}), all the objects plotted in Fig.~\ref{fig:vel}
definitely qualify for the HVG class introduced by Benetti et
al. (\cite{benetti05}).

From a photometric point of view, SN~2005ke has shown the same
properties as other 91bg-like objects (see Contreras et
al. \cite{contreras}), i.e. the absence of a secondary peak in the red
passbands, an intrinsically red color and a fast decline rate $\Delta
m_{15}(B)$=1.76$\pm$0.01 (Folatelli et al. \cite{folatelli}). For
comparison, SN~1999by and SN~2005bl reached $\Delta
m_{15}(B)$=1.90$\pm$0.05 (Garnavich et al. \cite{garnavich04}) and
1.93$\pm$0.10 (Taubenberger et al. \cite{tauben08}), respectively.
SN~2005ke reached a peak magnitude of $V$=14.2 (Contreras et
al. \cite{contreras}), which coupled to the estimated distance modulus
$\mu$=32.06$\pm$0.27 (Wood-Vasey et al. \cite{woodvasey}) yields an
absolute magnitude $M_V$=$-$17.9$\pm$0.3. We note that a fainter
magnitude is obtained ($M_V\sim-$17 ) if one uses the Tully-Fisher
distance modulus ($\mu$=31.16, Tully \cite{tully}).

\section{\label{sec:specpol}Spectropolarimetry}

Although we obtained two separate pre-maximum epochs (days $-$8.6 and
$-$7.6), a close analysis of the data shows that there is no
statistically significant evolution between the two dates. Therefore,
in view of the subluminous nature of the object, we have merged the
two data sets in order to maximize the signal-to-noise ratio. The
result is shown in Fig.~\ref{fig:dominant}, where we plot the combined
spectropolarimetric data set (corresponding to a total exposure time
of 1200 s per half-wave plate angle) on the $Q-U$ plane.

The data show a dominant axis (Wang et al. \cite{wang01}; Wang \&
Wheeler \cite{WW08}), not passing through the origin of the $Q-U$
plane, hence indicating the presence of different components,
characterized by different polarization degree and position angle (see
for instance Patat et al. \cite{patat09b}, their Appendix B). The
polarization variation along the dominant direction appears to be
moderate, but significant, reaching about 0.8\%. In contrast, the
polarization variation in the perpendicular direction is consistent
with the measurement errors (see Sect.~\ref{sec:pol}).

The first step for disentangling the different contributions is the
subtraction of any interstellar polarization (ISP) that might arise
within aligned asymmetric dust grains along the line of sight.

\subsection{\label{sec:isp}Reddening and interstellar polarization}

The Galactic reddening along the line of sight to SN~2005ke is
$E_{B-V}$=0.023 (Schlegel, Finkbeiner \& Davis \cite{schlegel}).
Using the relation found by Serkowski, Matheson \& Ford
(\cite{serkowski}) this implies that the Galactic interstellar
polarization (ISP) is expected to be lower than $\sim$0.2\%.  The
compilation of polarimetric data for Galactic stars by Heiles
(\cite{heiles}) contains only one entry within 3 degrees from
NGC~1371\footnote{http://vizier.cfa.harvard.edu/viz-bin/VizieR?-source=II/226},
i.e. HD~22332.  This star, at a projected distance of 1.1 degrees from
SN~2005ke, has a linear polarization $P$=0.13\% at position angle
PA=4.1 degrees (Matheson \& Ford \cite{matheson}), in good agreement
with the low Galactic extinction along this line of sight. Since the
extinction within the host galaxy is estimated to be comparably low
($E_{B-V}$=0.036$\pm$0.005; Folatelli et al. \cite{folatelli}), the
total ISP is expected to be of order of a few 0.1\%.

Several methods to derive the level of ISP in SN spectra have been
proposed in the literature (see for instance Leonard et
al. \cite{leonard00}; Howell et al. \cite{howell01}; Wang et
al. \cite{wang01}), and used in a number of cases in the SN field.
They are based either on the analysis of the $Q-U$ plane, or on the
assumption that some portions of the SN spectra are intrinsically
unpolarized (Wang, Wheeler \& H\"oflich \cite{wang97}, Howell et
al. \cite{howell01}, Chornock et al. \cite{chornock06}), so that a
non-null continuum polarization can be interpreted as arising within
the interstellar medium.

Although the application of these methods is possible for this object
(see below), we attempted an alternative solution, which is trying to
actually measure the ISP.  For this purpose we have obtained
spectropolarimetry of SN~2005ke at a rather late epoch, about two and
a half months past maximum light, when the SN is entering the
nebular phase. Since at these epochs the continuum polarization from
an asymmetric photosphere is expected to be null, any residual
continuum polarization can be interpreted as arising in intervening
interstellar dust. The very well-studied case of SN~2006X (Patat et
al. \cite{patat09a}) showed that, with only the remarkable exception
of the \ion{Ca}{ii} triplet, line polarization is null at epochs later
than one month past maximum light. This allows one to use a wide
wavelength range in order to reduce the uncertainty.

\begin{figure}
\centering
\includegraphics[width=9cm]{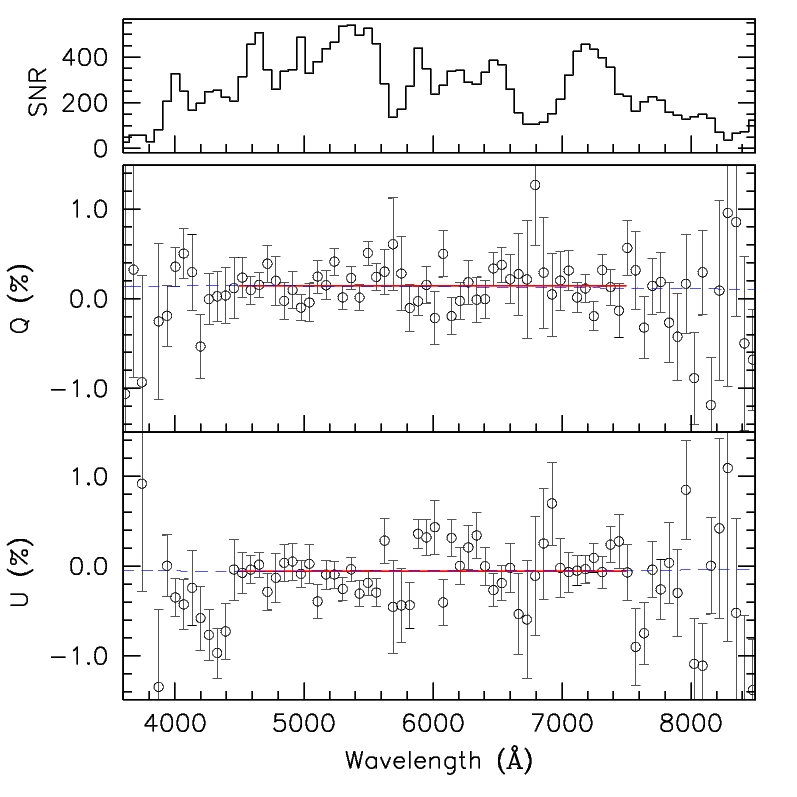}
\caption{\label{fig:quisp} Stokes parameters $Q$ (central panel) and
  $U$ (lower panel) on day +76. For presentation, the data were binned
  to 65 \AA\/ (25 px). The two horizontal segments mark the weighted
  average values within the wavelength range (4500-7500 \AA). The
  dashed curves trace a Serkowski-Whittet law with $P_{max}$=0.15\%,
  $\theta$=169.7 degrees, and $\lambda_{max}$=4500 \AA. The upper
  panel presents the signal-to-noise ratio per resolution element in
  the intensity spectrum.}
\end{figure}

\begin{table}
\centering
\tabcolsep 1.9mm
\caption{\label{tab:isp} ISP determinations}
\begin{tabular}{ccccc}
\hline
Range & phase & ID & $\langle Q_{isp}\rangle$ & $\langle U_{isp}\rangle$ \\
(\AA) &       &    & (\%)     & (\%) \\
\hline 
4800-5600 & pre-max & A& +0.007$\pm$0.012 & $-$0.186$\pm$0.017 \\
4000-4200 & pre-max & B& +0.155$\pm$0.040 & $-$0.053$\pm$0.047 \\
5100-5300 & pre-max & C& $-$0.029$\pm$0.025 & $-$0.303$\pm$0.030\\
\hline
4500-7500 & late    & D& +0.143$\pm$0.029 & $-$0.054$\pm$0.029\\
\hline
\end{tabular}
\end{table}

The spectropolarimetry of SN~2005ke on day +76 is presented in
Fig.~\ref{fig:quisp}. The data do not show any statistically significant
polarization variation across the whole spectral range.  The weighted
average Stokes parameters, estimated in the wavelength range 4500-7500
\AA, are $\langle Q_{ISP}\rangle$=0.14$\pm$0.03\% and $\langle
U_{ISP}\rangle$=$-$0.05$\pm$0.03\%, corresponding to a polarization
degree $P_{ISP}$=0.15$\pm$0.03\%, and a position angle
$\theta_{ISP}$=170$\pm$6 degrees. The rms deviation of single data points
from the weighted average value is 0.26\%, i.e. fully compatible
with the measurement errors. Because of the weakness of the SN at this
epoch ($V\sim$17.7, i.e. almost 3 magnitudes fainter than at maximum),
it is not possible to deduce the wavelength dependency of the
ISP. For a Serkowski-Whittet law (Serkowski et al.
\cite{serkowski}; Whittet et al. \cite{whittet}) the polarization
variation is very mild across the optical domain, and its absolute
amount is proportional to the maximum polarization. For the present
case, this implies a very small ($\leq$0.05\%) peak-to-peak variation
across the spectral domain covered by our observations, certainly
below the accuracy one can achieve with instruments like FORS (Patat \&
Romaniello \cite{patat06}).

In addition, we note that the ISP wavelength dependency observed in
several supernovae clearly deviates from the Galactic Serkowski law
(Leonard \& Filippenko \cite{leonard01}; Leonard et al.
\cite{leonard02}; Maund et al. \cite{maund07a}; Patat et
al. \cite{patat09a,patat09b}; Maund et al. \cite{maund10}). Therefore,
rather than using a Serkowski law to model the ISP, we ignored its
wavelength dependency and vectorially subtracted the constant values
derived above.  Once placed on the $Q-U$ plane, the derived ISP
happens to be very close to the dominant axis derived fitting all the
data of the combined spectropolarimetry obtained in the pre-maximum
epoch (Fig.~\ref{fig:dominant}, large filled circle). We also notice
that the measured ISP is very close to one of the two possible
solutions one would consider in the absence of late-time data, at
either end of the data along the dominant axis (see, for instance, the
ISP derivation for SN~1999by, Howell et al. \cite{howell01}). More
precisely, the measured ISP essentially coincides with the solution
implying the lowest ISP, which is fully in line with the low reddening
suffered by the SN.

\begin{figure*}
\centering
\includegraphics[width=13.5cm,angle=-90]{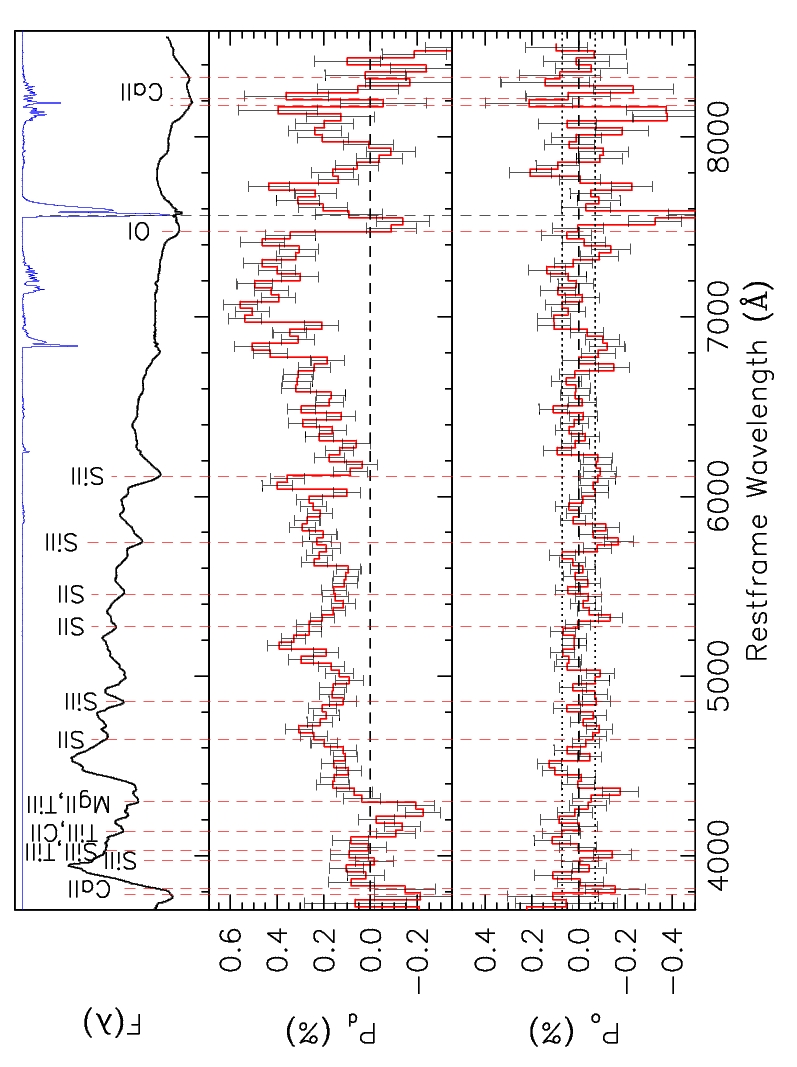}
\caption{\label{fig:rotated} Decomposition of polarization in the
  combined pre-maximum data set along the dominant (central panel) and
  orthogonal (lower panel) axis. The upper panel shows the un-binned
  flux spectrum and main line identifications; the thin curve traces
  a synthetic telluric absorption spectrum. The horizontal dotted
  lines in the lower panel indicate the rms deviation from null
  polarization.}
\end{figure*}

As an independent check, we have derived the ISP measuring the
polarization in the wavelength ranges 4800-5600 \AA\/ (A), 4000-4200
\AA\/ (B), and 5100-5300 \AA\/ (C) on the combined pre-maximum
spectrum. These ranges were proposed by Howell et al. (\cite{howell01}),
Chornock et al. (\cite{chornock06}) and Maund et al. (\cite{maund10})
respectively, based on the assumption that in these regions the line
blanketing is sufficient to produce complete depolarization. The
results are compared to those obtained from the late-time spectrum in
Table~\ref{tab:isp} and Fig.~\ref{fig:dominant} (circled crosses). The
ISP derived from the three wavelength ranges is $\leq$0.3\%, which is
consistent with the low level of polarization expected from the mild
extinction along the line of sight to SN~2005ke. The estimate obtained
for the B range coincides (within the statistical errors) with the
late-time measurement, indicating that in this particular spectral
region the depolarization assumption holds.

This is not the case (at least for SN~2005ke) for the other two
wavelength intervals, which give ISP values that are inconsistent with
the late-time data. In addition, the Stokes parameters derived from
range A place the ISP very close to the centroid of the data set (see
next section), so that notwithstanding the clear existence of a
dominant axis, the implied polarization angle would become erratic
(this would also turn into a very low intrinsic polarization level).
As for range C, it must be noticed that this was proposed for the
peculiar SN Ia 2005hk, where this particular region was indeed
expected to have null intrinsic polarization (Maund et al. \cite
{maund10}). This clearly does not apply to SN~2005ke.

We will adopt the ISP derived from the late-time data throughout the
paper.

\begin{table}
\centering
\caption{\label{tab:dominant} Best fit dominant axis parameters}
\tabcolsep 3mm
\begin{tabular}{lcclc}
\multicolumn{5}{c}{\bf Least Squares Fit Analysis}\\
\hline
$\alpha$ & $-$0.19\% (0.01\%) & & $\beta$  & 1.10 (0.13) \\
$\chi^2/DOF$ & 140/85 & & $r_{xy}$ & 0.43\\
$\sigma$ & 0.16\% & & $\theta_d$ & 113$^\circ$.9 (1$^\circ$.6)\\
\hline
 & & & & \\
\multicolumn{5}{c}{\bf Principal Components Analysis}\\
\hline
$\langle Q\rangle$ & 0.00 (0.01) &  & $\langle U \rangle$ & $-$0.20 (0.01)\\
b/a & 0.47 & & $\theta_d$ & 114.0 \\
\hline
\multicolumn{5}{l}{Note: values in parentheses indicate rms uncertainties.}
\end{tabular}
\end{table}

Since dust grains tend to be aligned along the direction of the
galactic magnetic field, which in turn follows the spiral pattern (see
Scarrot, Ward-Thompson \& Warren-Smith \cite{scarrot}), the ISP is
expected to be tangential to the local spiral arm of the host
galaxy. This has been observed, for instance, in SN~2001el (Wang et
al. \cite{wang03}), SN~2006X (Patat et al. \cite{patat09a}), in the
bright optical transient in NGC~300 (Patat et al. \cite{patat09b}),
and in SN~2005hk (Maund et al. \cite{maund10}). When the ISP within
the host galaxy and the Milky Way are comparable, as is the case for
SN~2005ke, the position angle of the resulting polarization vector can
differ substantially from the direction defined by the local spiral
pattern. This is the likely explaination why the derived ISP position angle
is far from being parallel to the spiral arm of NGC~1371 at the
location of SN~2005ke (Fig.~\ref{fig:fchart}). The ISP position angle
(170$\pm$6 degrees) is not too different from that reported for the
nearby star HD~22332 (4.1 degrees; see Sect.~\ref{sec:isp}). This
indicates that the Milky Way contribution may dominate over that
arising within the host.

\subsection{\label{sec:pol}Intrinsic polarization of SN~2005ke}

In order to derive the fundamental parameters of the dominant axis and
their statistical significance, we have performed a weighted linear
squares fit to the data (accounting for errors in both $Q$ and $U$) in
the wavelength range 4000-7500 \AA, using the linear relation
$U=\alpha\; + \beta\; Q$. The results are presented in
Table~\ref{tab:dominant}, where we included $\chi^2$ per degree of
freedom (DOF), the Pearsons correlation coefficient ($r_{xy}$), the
rms deviation of data points from the best fit axis ($\sigma)$, and
the dominant axis position angle ($\theta_d$).

In addition to the least squares analysis, we have performed a
principal components analysis. This was achieved constructing the
weighted covariance matrix and extracting its eigenvectors and
eigenvalues, as described in Maund et al. (\cite{maund10}). The
parameters retrieved through this analysis are the position angle of the
dominant axis ($\theta_d$) and the ratio ($b/a$) of the degrees of
polarization along the dominant and perpendicular axes. The results
are presented in Table~\ref{tab:dominant}, which includes the centroid
coordinates ($\langle Q \rangle$, $\langle U \rangle$). 

Following Wang et al. (\cite{wang03}), we computed the components of
the ISP-subtracted polarization along ($P_d$) and perpendicular
($P_o$) to the dominant axis, which are obtained by rotating
counterclockwise the observed $Q-U$ coordinates system by the angle
$\theta_d$ that defines the dominant axis (106.1 degrees). The result
is presented in Fig.~\ref{fig:rotated}, where we have also plotted the
main line identifications, following Taubenberger et
al. (\cite{tauben08}).

The first fact that emerges from this decomposition is that the bulk
of the polarization is aligned along the dominant axis.  The orthogonal
component $P_o$ is statistically consistent with a null value across
the whole wavelength range. One possible exception is seen at the
wavelength corresponding to the absorption trough of the \ion{O}{i} line at
about 7570 \AA. This deviation slightly exceeds the 3-sigma level
  ($P_o\sim$$-$0.50$\pm$0.15\%), and it takes place in a region which
  is partially affected by a strong telluric feature (see the telluric
  absorption spectrum in Fig.~\ref{fig:rotated}, upper panel), and
  therefore it needs to be considered with caution. Taken at face
  value, this would imply that oxygen is polarized along a direction
  orthogonal to that displayed by the underlying continuum. 

\begin{figure}
\centering
\includegraphics[width=8.0cm]{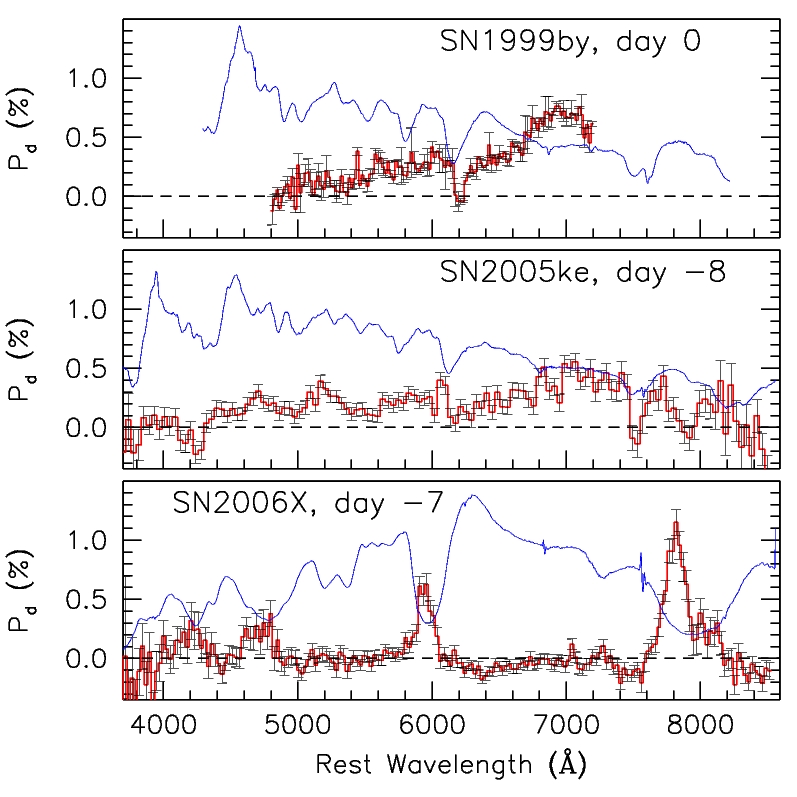}
\caption{Comparison between the dominant axis polarization spectra of
  SNe 1999by (top; Howell et al. \cite{howell01}), 2005ke
  (middle), and 2006X (bottom; Patat et al. \cite{patat09a}). The
    dominant component for SN~1999by was computed using $\theta_d$=80
    degrees (Howell et al. \cite{howell01}). For presentation, the
    original 1999by data were binned in $\sim$40 \AA\/ wide bins (11
    pixels). The dashed horizontal line marks the zero-polarization level.
    The thin (blue) lines trace the observed flux
  spectra. No reddening correction was applied to SN~2006X.}
\label{fig:compol}
\end{figure}

  The second remarkable property revealed in Fig.~\ref{fig:rotated} 
  is the relatively low polarizations
  corresponding to the main absorption features such as \ion{Si}{ii}
  $\lambda$6355 and the NIR \ion{Ca}{ii} triplet, with respect to that
  typically observed in Type Ia SN a week before maximum light (Wang
  \& Wheeler \cite{WW08}). Although there are some localized, weak
  peaks in $P_d$ (some of which are statistically significant), the
  polarization is always below $\sim$0.5\%. The polarization is dominated by the
  continuum, which shows an increase from blue to red peaking at about
  7200 \AA. This is reminiscent of what was found for SN~1999by
  (Howell et al. \cite{howell01}), and establishes a clear link
  between these two subluminous events (see Fig.~\ref{fig:compol}, top
  and middle panels). In addition, the continuum polarization marks a clear
  difference with respect to core-normal events. In core-normal
  events, the continuum polarization in the range 6400-7000 \AA\/ is
  always below 0.2\% (see for instance SN~2006X, Patat et
  al. \cite{patat09a}). The overall differences between the
    subluminous events (SN~1999by, SN~2005ke) and the core-normal
    SN~2006X are illustrated in Fig.~\ref{fig:compol}, which presents
    the polarization spectra along the respective dominant axes.

\begin{figure}
\centering
\includegraphics[width=8cm]{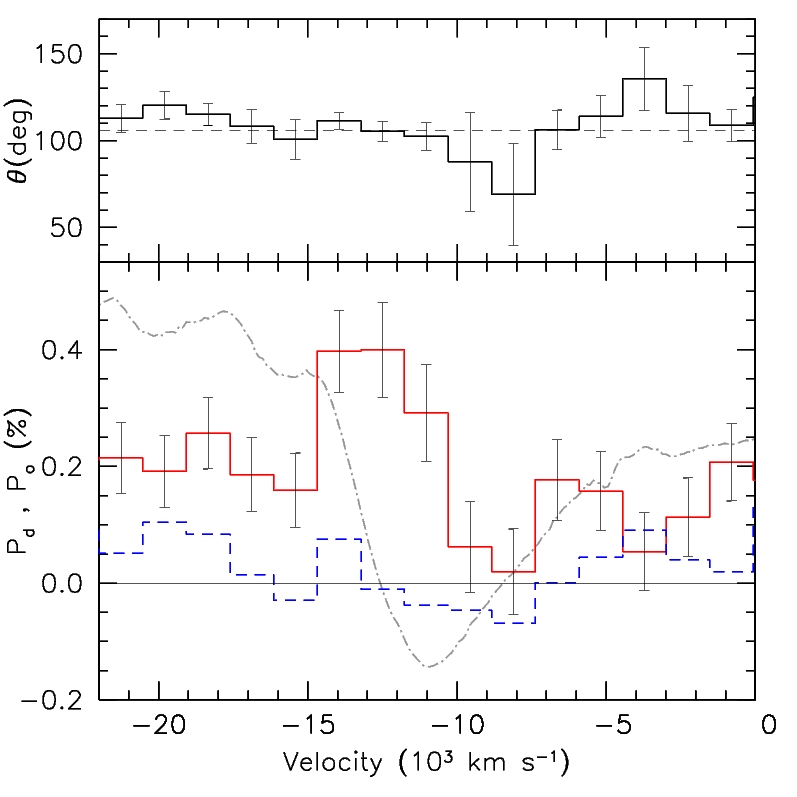}
\caption{\label{fig:siprof}Lower panel: polarization profile of
  \ion{Si}{ii} $\lambda$6355 at the combined pre-maximum epoch. Solid
  and dashed lines trace $P_d$ and $P_o$, respectively (12 px bins,
  $\sim$35 \AA).  The dotted-dashed curve is the unbinned flux
  spectrum arbitrarily scaled. Upper panel: polarization position
  angle. The horizontal dashed line indicates the dominant axis angle
  $\theta_d$.}
\end{figure}

The most marked feature in the dominant-axis component is the peak
corresponding to the \ion{Si}{ii} $\lambda$6355 absorption. The
polarization profile, decomposed along and orthogonally to the
dominant axis is shown in Fig.~\ref{fig:siprof}. The polarization peak
($P_d\sim$0.39$\pm$0.08\%) occurs at $-$13,300 km s$^{-1}$, while the
minimum of the absorption trough is at a lower velocity ($-$10,800 km
s$^{-1}$), as is typical of Type Ia SNe (see for instance the well
studied case of SN~2006X; Patat et al. \cite{patat09a}). The
polarization angle across the line is constant (within the errors),
and equal to $\theta_d$, suggesting that the line forming region and
the photosphere share the same geometrical properties.  The
signal-to-noise ratio is insufficient to tell whether the polarization
angle is constant across the NIR \ion{Ca}{ii} triplet as well.

\section{\label{sec:models}Models for SN~2005ke}

In this section, we compare the observations of SN~2005ke with
possible theoretical scenarios.  As in previous studies, the results
have been obtained using our hydrodynamical radiation code, HYDRA.
This allows us to solve the hydrodynamics including nuclear reaction
networks for the explosion and radiation transport, including detailed
atomic models for light curves, flux and polarization spectra.  The
methods are described in H\"oflich (\cite{h91, h95}), Howell et
al. (\cite{howell01}), H\"oflich et al. (\cite{h02, h06, h09}), Wang
et al. (\cite{wang06b}).  For calculating the polarization spectra in
our Monte Carlo module, we use the given density, temperature and
chemical structures including the non-LTE occupation of the atomic
models.  We can do so because polarization will not affect the rate
equations within the accuracy of our approximations.  In SNe, most of
the intrinsic polarization is due to Thomson scattering; nevertheless,
our calculations take into account also Rayleigh scattering in lines,
under the assumption of complete redistribution. Thus, the scattering
matrix is approximated by a linear combination of Rayleigh and
isotropic shift in phase (Hamilton \cite{hamilton1947}; Domke \&
Hubeny \cite{domke88}) with the size of polarization given by the
angular momentum of the lower atomic state (Chandrasekhar
\cite{chandrasekhar60}).

In the first part of this section, we will discuss the origin of the
polarization detected in SN~2005ke. This is done in the framework of a
subluminous delayed-detonation (DD) model of a Chandrasekhar mass
White Dwarf (WD) because many properties of SN1991bg-like events can
be understood within this family of models.  We note that the
deflagration-detonation transition may occur both during the
initial expansion phase or after a pulsation (Khokhlov, M\"uller \&
H\"oflich \cite{khokhlov93}; H\"oflich \cite{h95}; H\"oflich et
al. \cite{h02}). We choose a scenario with $M_{WD}$=$M_{Ch}$ because
only these models burn under sufficiently high densities to undergo
electron capture as indicated by observations (H\"oflich et al. \cite
{h04}; Motohara et al. \cite{motohara06}; Fesen et al. \cite{fesen07};
Maeda et al. \cite{maeda11}). We note that the evidence for electron
capture has been found in core-normal and moderately subluminous
SNe~Ia, SN~1986G-like events, but there is a lack of observations of
91bg-like events.  
Many of the arguments presented here are generic and they are also used to
discuss possible alternative scenarios including mergers and
edge-lit sub-Chandrasekhar mass models.  

\subsection{\label{sec:ddt}Delayed Detonation Model} 

We start with a spherical model and remap that structure to
aspherical grids. We can do so because the deviations from sphericity
are found to be small. The remapping allows us to study various
effects, such as rotation. These effects may lead to oblate, ellipsoidal
structures (Eriguchi \cite{eriguchi85}) and off-center Ni
distributions, as may be produced by off-center DD (Livne
\cite{livne99}; Gamezo, Khokhlov \& Oran \cite{gamezo05}; H\"oflich \&
Khokhlov \cite{h06}; R\"opke \cite{roepke07}; Kasen et
al. \cite{kasen09}).

Within the DD scenario, the free parameters and prescriptions are: 1) the chemical
structure of the exploding WD, 2) the central density, $\rho_c$, at
the time of the explosion, 3) the description of the deflagration
front, and 4) the density, $\rho_{tr}$, at which the transition from
deflagration to detonation occurs.  In this work, the spherical
structure of the exploding C-O WD is based on a solar metallicity
model star with 5 M$_{\sun}$ at the main sequence.  Through accretion,
this core has been grown close to the Chandrasekhar limit (see model
5p0y23z22 in Dominguez et al. \cite{dominguez01}).  At the time of the
explosion of the WD, its central density $\rho_c$ is 2.0$\times 10^9$
g~cm$^{-3}$ and its mass is close to 1.37$M_\odot$. The transition
density $\rho_{tr}$ has been identified as the main factor that
determines the $^{56}$Ni production, and thus the brightness of a
SNe~Ia (H\"oflich \cite{h95}; H\"oflich, Khokhlov \& Wheeler
\cite{hkw95}; Iwamoto et al. \cite{iwamoto99}). The
deflagration-to-detonation transition density is $\rho_{tr}$=8$\times
10^6$ g~cm$^{-3}$. This value, which is lower than that typical
  of core-normal models (2$\times$10$^7$ g~cm$^{-3}$), is required to
  produce a subluminous explosion (H\"oflich et al. \cite{h02}).  We
employ this model (5p0z22.8, H\"oflich et al. \cite{h02}), because it
reproduces the light curves and spectra of SN~1999by from the optical
to the NIR. We note that, although some $0.3 M_\odot$ are burned
during the slow deflagration phase, only $\sim0.09 M_\odot$ of
$^{56}$Ni are produced because of electron capture. The light curve
reaches its maximum about 14.6 days after the explosion, at
$M_V\sim-$17.2.

\begin{figure}
\centering
\includegraphics[width=9cm]{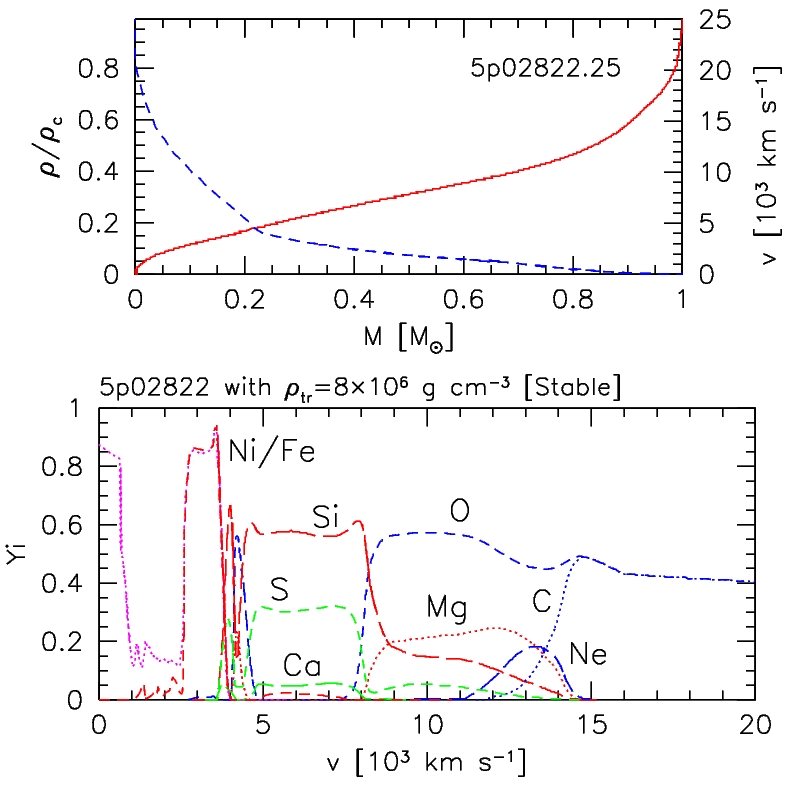}
\caption{Density (dotted) and velocity (solid) as a function of
  enclosed mass (upper panel), and abundances of stable isotopes
  (lower panel) as a function of the expansion velocity for DD-model
  5p0z22.8, with a transition density of $\rho_{tr}=$ 8 $\times 10^6$
  g cm$^{-3}$. In addition, $^{56}$Ni is given (from H\"oflich et al. \cite{h02}).}
\label{fig:abundance}
\end{figure}

The final density, velocity and chemical structures are given in
Fig. \ref{fig:abundance}. The density and velocity profiles are very
similar to core-normal SNe~Ia, because the specific nuclear energy is
nearly the same when burning from C-O to Si- or Fe-group elements;
however, more partially burned or unburned material is left in the
subluminous models.  Due to the larger pre-expansion during the
deflagration phase, most of the star is burned to Si-group elements
rather than to $^{56}$Ni. This structure is consistent with those
obtained by inverse spectral analysis (Mazzali et
al. \cite{mazzali97}). A further feature that distinguishes
subluminous models from core-normal DD models is the extended region
of unburned C-O.  As discussed earlier (see also
Fig.~\ref{fig:siprof}), the Si line profile extends in the blue up to
about 16,000 km s$^{-1}$, consistent with the chemical structure of
this model. The photosphere is at a velocity of about 10,000 km
s$^{-1}$, i.e. well within the Si-rich zone.

\subsection{\label{sec:flux}Flux spectra} 

\begin{figure*}
\centering
\includegraphics[width=18cm]{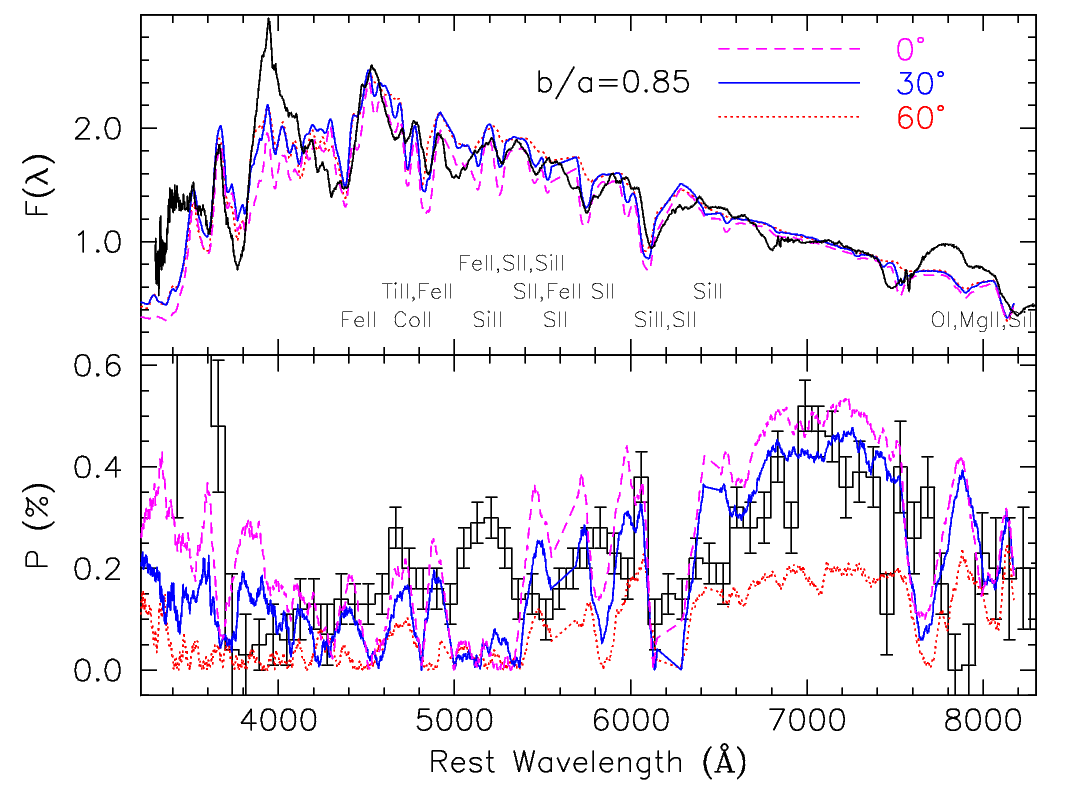}
\caption{Model flux (top) and polarization (bottom) spectra of the
  subluminous DD model 5p0z22.8 at 7 days before maximum light at
  three inclination angles in comparison with the data of SN~2005ke ($\sim$8
  days before maximum). We assumed an aspherical density structure
  with an axis ratio b/a=0.82.  We show the observations (black, with
  error bars for the polarization). For presentation, the observed polarization 
  data were binned to 78 \AA\/ (30 pix).}
\label{fig:spec}
\end{figure*}

In Fig. \ref{fig:spec}, we compare spectra of SN~2005ke at 8 days
before maximum with theoretical spectra at about day 9 after the
explosion assuming an ellipsoidal geometry with an axis ratio
b/a=0.85. 

The model is taken directly from the time evolution of the initial
explosion model without further tuning.  The adopted b/a axis ratio
allows us to reproduce the size of the polarization in the red for
inclinations between 30 and 60 degrees.

Overall, the theoretical spectra provide a good match to the observed
flux spectra of SN~2005ke: the overall energy distribution, the
presence and Doppler shifts of the lines are well reproduced.  The
spectrum is dominated by single ionized stages of Si, S, Ti and the
iron group elements. Given the simplifying assumptions made,
one cannot expect a perfect fit. For one, we assumed that the time
evolution is identical to the spherical case, and we calculated the
asymmetric profiles for a specific point in time. We had to do so
because detailed atomic models for the ions are beyond our current
computational means. This, however, is a good first-order
approximation, again because the deviations from sphericity are small. The
theoretical line strengths are slightly larger for most inclinations,
which may indicate a larger than predicted population
of excited levels of iron-group elements. In addition, the
models have more line blocking at about 5200 \AA, and the \ion{Ti}{ii}
feature at about 4800 \AA\/ is weaker than observed, as is the \ion{Ca}{ii} emission
at about 3800 \AA.  From previous systematic studies of the
temperature effect on spectra (Nugent et al. \cite{nugent95}), we
estimate that the local temperature in our model is too high by about
1000 K.

\subsection{\label{sec:oblate}Polarization spectra for oblate geometries} 

The overall polarization agrees with the data of SN~2005ke
(Fig. \ref{fig:spec}, lower panel). In the optical, P increases with
increasing wavelength, and shows a maximum around 7000 \AA. The
largest discrepancy between the models and the data is seen at about
5200 \AA, where the calculations predict little or no polarization,
while the observations show $P \approx 0.2 $ to $ 0.3 \% $, i.e.  well
outside the error bars. A missing feature is the depolarization at
about 8000 \AA\/ which may be attributed to shell interaction with the
environment (Gerardy et al. \cite{gerardy}, Quimby et al. \cite{q06}),
but has not been included in this model. To fit such a feature, two
additional parameters would have to be added, the orientation and the
mass of the circumstellar matter.  Note that the seemingly high
  observed polarization below 3600 \AA\/ is affected by very large
  uncertainties due to the low flux levels.
 
In the following, we will discuss the polarization results.  In
general, polarization provides information about asphericity on
scales comparable to the mean free path of photons. Polarization is
decreased if the radiation field becomes more isotropic, e.g. by
thermalization (H\"oflich \cite{h91}).

The origin of line and continuum polarization in rapidly expanding
atmospheres dominated by Thomson scattering can be represented by
three idealized scenarios (H\"oflich \cite{h91}; H\"oflich \cite{h95};
Kasen et al. \cite{kasen03}). These are: 1) aspherical density
distributions e.g. by rapid rotation of a single WD or binary WD mergers, 2)
anisotropy in the radiation field, e.g. aspherical $^{56}$Ni
distributions, or 3) selective line blocking of the underlying
``photo-disk'', e.g. by aspherical chemical distributions as caused by
off-center DDs. We first consider the latter mechanism. The alternative polarization
mechanism will be considered in Sections \ref{sec:prolate} and \ref{sec:sub}.

\subsubsection{\label{sec:lines}Line polarization}

Selective line blocking requires a non-radial chemical gradient in the
line forming region and can occur even though the Thomson scattering
photosphere may be spherical. As a result, the "photo-disk" shows
local polarization increasing with the distance from the center,
because perpendicularly scattered photons and forward scattered
photons are 100\% and 0\% polarized, respectively (note that, at the
photosphere, not all photons are traveling radially even at the rim of
a disk). For a sphere, the components of the polarization field cancel
out. If the chemical distribution is asymmetric, the disk may be only
partially obscured at the wavelength of the line.  If seen off-axis,
the line may be optically thick in one part of the disk and optically
thin in another. Thus, the polarization vectors of the underlying
Thomson photosphere incompletely cancel yielding a high line 
polarization. This effect can cause a line
polarization of the order of 2\% if the line-forming region is
detached from the photosphere, or for strong lines in rapidly
expanding photospheres of SNe.


Strong polarization in lines is produced by '"asymmetric covering" of the 
underlying Thomson scattering-dominated photosphere by asymmetries
in the abundance distribution. For strong lines, the amount of material
required to selectively block the photosphere may be small.
Line polarization is a transient phenomenon, dictated by the position
of the receding photosphere with respect to the boundaries of chemical
asymmetries (see also H\"oflich et al. \cite{h06} and the discussion in Patat et
al. \cite{patat09a}).  High line polarization levels are commonly seen in
\ion{Ca}{ii}, \ion{Mg}{ii}, \ion{S}{ii} and \ion{Si}{ii} in many
core-normal SNe at early times (Wang et al. \cite{wang03}; Kasen et
al. \cite{kasen03}; Wang et al. \cite{wang04, wang06b}; Patat et
al. \cite {patat09a}; Maund et al. \cite{maund10}).  In these objects,
the effect is large when the photosphere is close to the chemical boundary, 
e.g. between the Mg/Ne/O and Si/S-rich layers.
In subluminous events,  the corresponding lines block the entire photosphere,
because the Thomson photosphere is smaller, and the lines are formed in an 
extended region of Si/S (see below).

Although the models included Thomson scattering on free electrons and
Rayleigh scattering in lines, the former is larger by a factor of 10
and, for the sake of the argument, we assume that absorption in a line
depolarizes the radiation field. In SN~2005ke and at $-$8 days,
  the lines are formed at about 10,000-12,000 km s$^{-1}$, i.e. far
  from chemical boundaries (see Fig.~\ref{fig:abundance}). Therefore,
  line polarization by selective line blocking is small for strong
  lines such as \ion{Mg}{ii}, \ion{Si}{ii} and \ion{S}{ii}.  Intrinsic
  line polarization due to Rayleigh scattering is small for most line
  transitions; however, high polarization is sometimes observed in
  strong lines if they form in regions of asymmetric chemical
  distributions. The resulting asymmetric blocking of the light from
  the scattering photosphere can result in a high net polarization. In
  the model for SN~2005ke, the large extension in velocity space of
  zones of C burning and incomplete O burning causes blocking of
  the entire photosphere, so strong lines forming in that region do not display
  polarization. Weak lines form close to the photosphere and, thus,
  always result in small polarization.  In conclusion, the
  polarization spectrum is dominated by depolarization in strong
  lines.

\begin{figure}
\centering
\includegraphics[width=9.0cm]{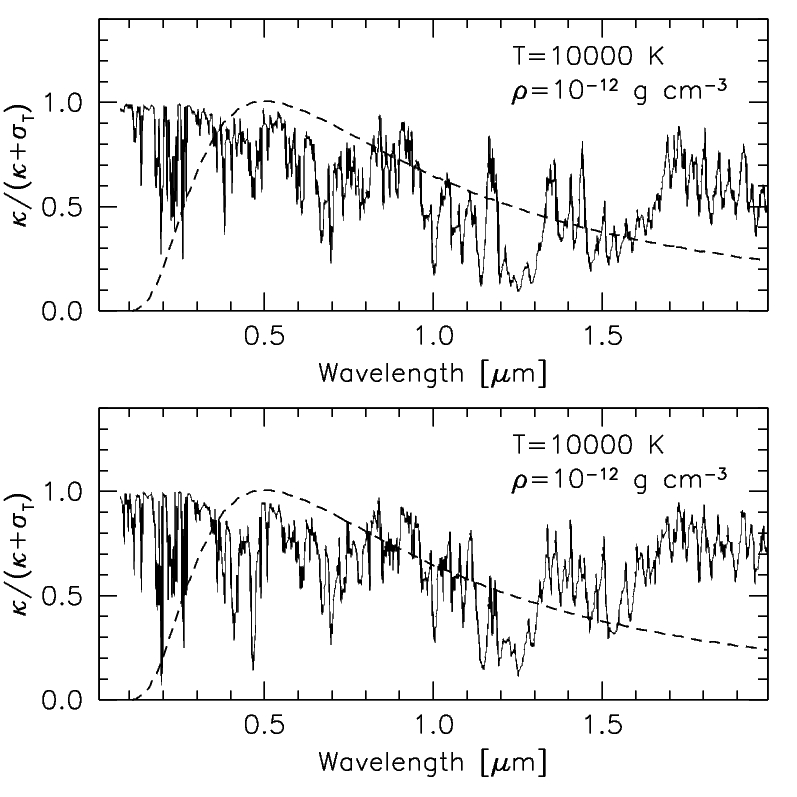}
\caption{Comparison of the ratio $R$ of line to total opacity for
  conditions typical for the photosphere of a SN Ia at the
  thermalization optical depth which corresponds to an optical depth
  of $\approx$2 to 3 in Thomson scattering for photons in the optical
  and near IR.  Line opacity is given at 10,000 K and for a density of
  $10^{-12}$g cm$^{-3}$ for the Si-rich (top) and Fe-rich (bottom)
  composition (from H\"oflich \cite{h93}), corresponding to the
  thermalization and last scattering layers in subluminous and
  core-normal models, respectively.  In the red (at $\approx 7000
  \AA$) the average $R$ is higher by about 50\% in the latter. The
  scattering optical depth at the last scattering radius is about
  $0.5$ and $0.3$ for subluminous and core-normal SNe~Ia,
  respectively.  As a consequence, continuum polarization will be
  smaller for the same degree of asphericity (see text). The dashed
  lines are the flux distributions for Planck functions of the
  corresponding effective temperature.}
\label{fig:opacities}
\end{figure}

\subsubsection{\label{sec:continuum}Continuum polarization}

In addition to the polarization for line features, it is also
important to understand the polarization behavior of the continuum. In
SNe Ia, a pseudo-continuum is formed by blending of a large number of
lines which are mostly responsible for thermalization (Karp
\cite{karp77}).  The wavelength dependence of the pseudo-continuum is
one of the keys to understand the wavelength dependence of $P$.  The
ratio $R$ between true absorption $\kappa$ and total extinction,
$\kappa + \sigma_{Thomson}$, determines the thermalization optical
depth, $\tau \approx \sqrt{3/R}$ and, more important, the ratio
between polarizing, i.e. Thomson, and non-polarizing extinction at
the radius of last interaction, $\tau_{extinction} \approx 1$.  In
Fig. \ref{fig:opacities}, $R$ is given for Si and Fe-rich mixtures for
conditions typical of Type Ia SN photospheres before maximum light.
Photons thermalize and are depolarized at Thomson optical depths
between 0.1 to 3 and 0.1 to 1, respectively.  The ratio
depends strongly on wavelength. In the UV, where $R$ is close to
unity, the opacities are line dominated whereas at longer
wavelengths,Thomson scattering becomes more important.

\begin{figure}
\centering \includegraphics[width=8.5 cm]{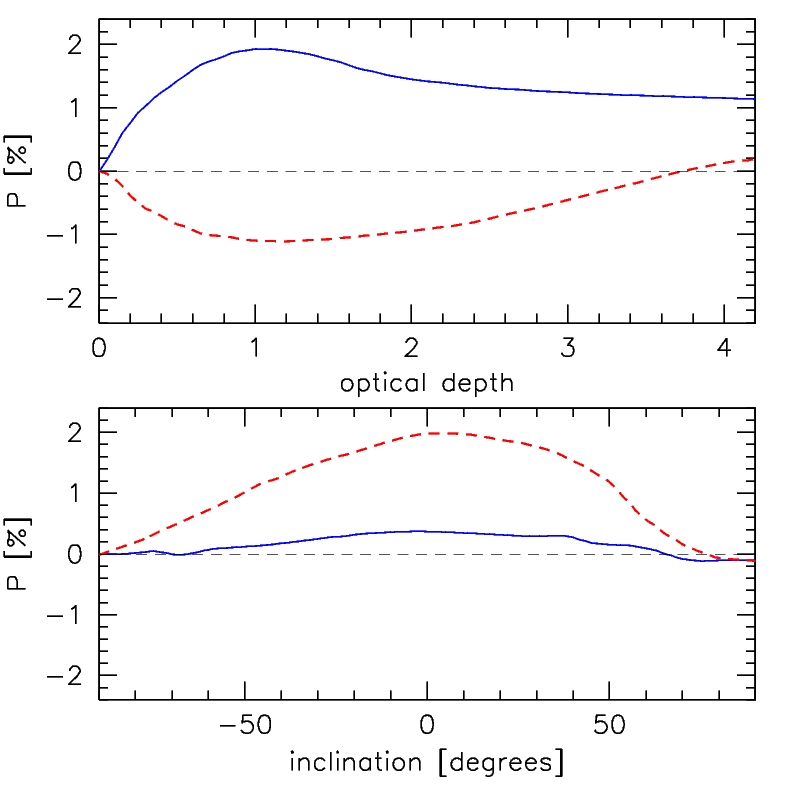}
\caption{\label{fig:p1} Continuum polarization $P$ for aspherical
  configurations in scattering dominated atmospheres with an electron
  distribution $\propto r^{-3}$.  Upper panel: maximum linear
  polarization as a function of the optical depth for oblate (solid
  blue) and prolate (dashed red) ellipsoids. Lower panel: $P$ as a
  function of the inclination for a source located off-center, at a
  distance D=$R_{ph}/R$ of 3 (solid blue) and 1.5 (dotted red)
  (adopted from H\"oflich et al. \cite{hkw95}).  Note that $D=3$
  corresponds to SN~2005ke at day $-$8, when the photosphere $R_{ph}$
  is formed at about 10,000 km s$^{-1}$, and assuming the maximum
  off-center $^{56}$Ni distribution suggested by Maeda et
  al. (\cite{maeda11}) for core-normal SNe~Ia.}
\end{figure}

A second key for the understanding of continuum polarization in
scattering-dominated atmospheres is the behavior of polarization as a
function of optical depth and degree of asymmetry (H\"oflich
\cite{h91}). In Fig.~\ref{fig:p1}, we show the continuum polarization
$P(\tau)$ for various configurations. Let us first consider an oblate
ellipsoid.  If the absorptive optical depth is small, $P$ increases
linearly with $\tau_{sc}$ because of the increasing likelihood that a
photon undergoes scattering.  It reaches a maximum at $\tau\sim$ 1.2
and, then, declines by about 60\%, as shown by the solid curve in
Fig.~\ref{fig:p1} (upper panel). This decrease in polarization at larger 
optical depth is because, with multiple scattering, the radiation field 
becomes more isotropic. Note that for an oblate geometry, $P$ varies 
with optical depth, but does not change sign, i.e.  the polarization angle 
is constant.
 
This has two important implications for our understanding of
polarization spectra: 1) the continuum polarization degree is not a
linear function of the opacity in the pseudo-continuum, and 2) the
continuum polarization angle may change with wavelength even for
axially symmetric configurations (i.e. $P $ may be not positive
definite) for prolate structures or off-center energy sources (see
Sect.~\ref{sec:prolate}).
For oblate ellipsoids the wavelength dependencies of opacity and
  continuum polarization are related. Therefore, the wavelength
  dependence of $P$ seen in SN~2005ke (Fig. \ref{fig:spec}) can be
  understood in terms of the variation of the optical depth with
  wavelength of the photosphere for a Si-rich chemistry, which is
  relevant to subluminous SNe~Ia (as opposed to Fe-rich, core-normal
  events). At short wavelengths, lines form a pseudo-continuum and
  dominate the opacity. As a consequence, $P$ is low. Going to longer
  wavelengths, on average\footnote{The curves shown in
    Fig.~\ref{fig:opacities}) are a realization at a single value of
    velocity, and do not include the blurring produced by the velocity
    gradient within the decoupling region. Although $R$ is sensitive
    to wavelength on small wavelength scales in
    Fig.~\ref{fig:opacities}), the photosphere forms over a range of
    velocities, leading to a significant velocity smearing in the
    effective value of $R$ in a given wavelength range.} $R$ decreases
  to attain a minimum at about 7000 \AA\/ ($R\approx 0.5$) and $P$
  reaches its maximum.

Between 5000 and 5400 \AA, blending depends sensitively on the
temperature, as already discussed in the context of the flux spectrum
(see Sect.~\ref{sec:flux}).  Therefore, the discrepancy between the
model and the data seen around 5200 \AA\/ (see Fig.~\ref{fig:spec},
lower panel) may be explained in terms of temperature changes, because
the energy differences between levels in heavy elements is comparable
to the thermal energy.

\vspace{2mm} In general, for oblate geometries, the polarization
spectrum of SN~2005ke can be explained in terms of an intrinsically
polarized continuum depolarized by lines.

\section{\label{sec:disc}Discussion}

\begin{figure}
\centering
\includegraphics[width=9cm]{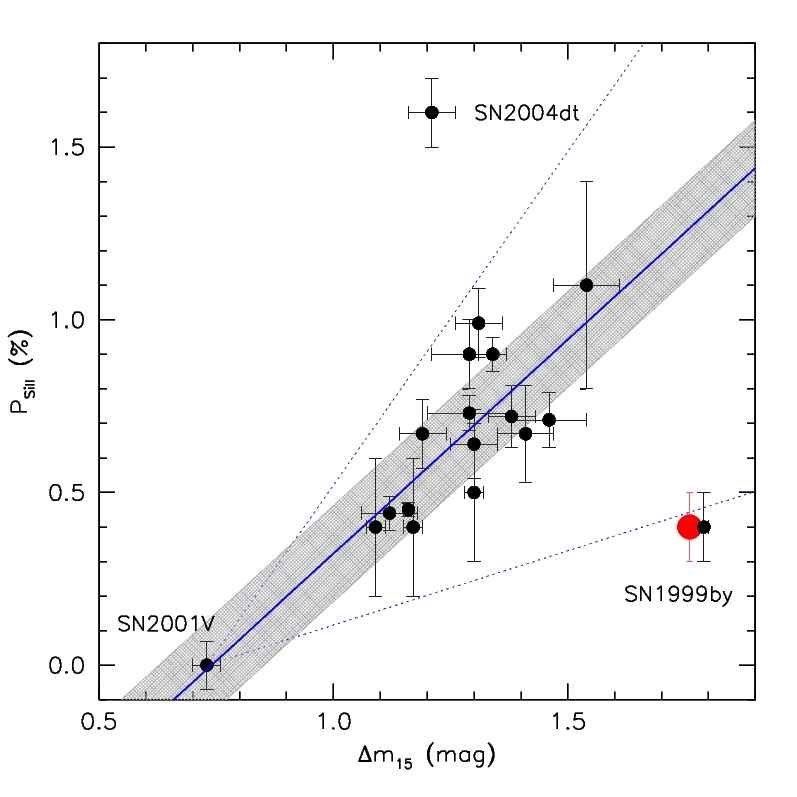}
\caption{\label{fig:poldm}Degree of polarization across the
  \ion{Si}{ii} $\lambda$6355 line as a function of light curve decline
  rate. Data are from Wang et al. (\cite{wang07}) and Patat et
  al. (\cite{patat09a}).  SN~2005ke is marked by the large dot close
  to SN~1999by.  The solid line is a best fit to all data, with the
  exception of SN~1999by, 2004dt, and 2005ke. The dotted lines trace
  the 1-$\sigma$ level of the intrinsic polarization distribution
  generated by the Monte-Carlo simulation discussed by Wang et
  al. (\cite{wang07}). The shaded area indicates the rms deviation of
  the data points from the best fit relation.}
\end{figure}

SN~2005ke is a close spectroscopic and photometric twin of SN~1999by,
and both show similar polarization properties, with respect to the
amplitude of the polarization ($\sim$0.6\%) as well as wavelength
dependency.  SNe~1999by and 2005ke have very distinct polarization
properties compared to core-normal events (see also
Fig.~\ref{fig:compol}), which place them in a fairly isolated region
of the $\Delta m_{15}$ - P plot (Fig.~\ref{fig:poldm}). They also
appear as outliers from the relation between the \ion{Si}{ii} velocity
gradient and the associated polarization (Maund et
al. \cite{maund10b}).

The difference in appearance of subluminous events is caused by low
photospheric temperatures in combination with layers of unburned
carbon and more massive layers of the products of explosive carbon and
oxygen burning.

The polarization of SN~2005ke can be understood in the framework of an
oblate ellipsoidal geometry, with a degree of asymmetry of about 15\%.
If this is typical of subluminous Type Ia events, then a spread in
the apparent peak brightness by about 0.2 mag is expected, just
because of the directional dependence of the luminosity.

One important question to ask is what causes the overall asymmetry in
these objects, as opposed to what is seen in core-normal SNe Ia
explosions like SN~2006X (Patat et al. \cite{patat09b}. See also
Fig.~\ref{fig:compol}), and whether this can tell us something about
the explosion mechanism that leads to these events.

It is tempting to attribute the asymmetry to the initial WD rotation,
as first proposed by Howell et al. (\cite{howell01}) for SN~1999by. As
a first approximation, we assume that the original degree of asymmetry
is conserved during the explosion. As for mass redistribution, three
dimensional calculations for SNe~Ia explosions show little coupling
between latitudes.  When starting from a spherical WD, the resulting
envelope remains spherical even if the burning is highly asymmetric
(e.g. Gamezo et al. \cite{gamezo05}).  At about 7 days after the
explosion, about 8 days before maximum light, the photosphere is
formed in a layer enclosing $\sim$0.7 $M_{WD}$
(Fig. \ref{fig:abundance}).  Based on the work on rotating WDs
(Eriguchi \& M\"uller \cite{eriguchi85, eriguchi93}; M\"uller \&
Eriguchi \cite{mueller85}), a rotation velocity close to the break-up
velocity produces an asphericity of about 15\% at layers placed at 0.7
in mass coordinates. Fast rotation may be reasonable because, within
the single degenerate scenario, mass is accreted from a companion star
via an accretion disk rotating at near-Keplerian velocities, although
this aspect does not differentiate subluminous SNe from single
degenerate models of core-normal events.

If the scenario in which the overall asymmetry is inherited from the
WD rotation is correct, then one can reverse the argument, and
speculate that a low (or null) polarization indicates a slow (or no)
WD rotation. Now, for core-normal SN~Ia, we have upper limits to the
continuum polarization only before and around maximum light. This
limit is 0.1 to 0.2\% (Wang et al. \cite{wang06b}; Wang \& Wheeler
\cite{WW08}, Patat et al. \cite{patat09a}). This indicates that fast
WD rotation and SN sub-luminosity may be related, marking a clear
distinction with respect to core-normal explosions.

At this point one can ask: do we expect a similar degree of
polarization in normal and subluminous events for the same degree of
asymmetry?  In core-normal Type Ia and around maximum light, the
thermalization of pseudo-continuum photons in the red takes place
mostly in layers of the iron group, at variance with objects like
SN~2005ke, in which thermalisation occurs in layers of incomplete burning.  The
ratio $R$ between true absorption and total extinction (see
Sect.~\ref{sec:oblate}), averaged over the region 6750-7250 \AA, is
0.7 and 0.5 in core-normal and subluminous supernovae,
respectively. There are two reasons for the difference in $R$. Firstly, the
thermalization layers changes from iron-rich to Si/S-rich compositions
(Sect.~\ref{sec:oblate}). Secondly, for core-normal SNe~Ia, the
thermalization happens at layers of larger expansion velocity and,
thus, smaller density leading to a further increase of $R$ (see
Hoeflich et al. \cite{h93}, their Fig.~1).

Moreover, larger thermalization in core-normal SNe~Ia results in an
increased isotropy of the radiation field (H\"oflich \cite{h91}).  For
the same amount of asymmetry, the intrinsic polarization at about 7000
\AA\/ may be smaller by up to a factor $\approx 2$ for core-normal
SNe~Ia (Figs. \ref{fig:opacities}), \ref{fig:p1}), but not a factor 5
to 6 as indicated by early-time spectropolarimetric data. Though not
conclusive, because of the small number of early-time observations for
SNe~Ia, there indeed may be a connection between rotation and
sub-luminosity. In this respect, it is interesting to note that a
relatively massive (1.28$\pm$0.05 M$_\odot$) WD, with a rapid rotation
(P=13.2s) has been found in a binary system (Mereghetti et
al. \cite{mereghetti11}). The companion star is a mass-loosing,
  hot sub-dwarf, from which the WD is currently accreting material.

\subsection{\label{sec:prolate}Prolate geometries and off-center energy sources}

Given the lack of spectropolarimetry time coverage, alternative
geometries within $M_{Ch}$ explosions need to be considered,
namely prolate ellipsoids and off-center energy sources.
These can be understood within the same context discussed in
Sect.~\ref{sec:oblate}. In prolate geometries, tangential rays
dominate the polarization for small optical depths causing negative
$P$.  With increasing optical depth, multiple scattering causes the
radiation field to become more isotropic. This reduces the tangential
flux and causes $P$ to change sign at greater optical depth
(Fig.\ref{fig:p1}, and H\"oflich \cite{h91,h95}).

In the light of these considerations, we first note that overall
prolate structures can be ruled out for SN~2005ke. The observed
polarization is positive at all wavelengths whereas, in the case of a
prolate geometry, it should change its sign (i.e. angle) at
wavelengths dominated by Thomson scattering (Fig. \ref{fig:p1}).

As for off-center energy sources (i.e. $^{56}$Ni) in Type Ia SNe,
these have been inferred from the NIR line profile at 1.65 $\mu$m
(Maeda et al. \cite{maeda11}), and from observations of the remnant of
S-Andromedae, which has been attributed to the class of SN~1986G-like
objects (Fesen et al \cite{fesen07}). The amount of "off-centerness"
is less than about 2000 km s$^{-1}$. The observations of remnants and 
NIR line asymmetries are limited to a few cases only, none of which is 
a SN~1991bg-like event. For DD models, a week before maximum, 
the optical depth of the central region is about 50 (H\"oflich \cite{h95}).  
At $-$8 days, the inner layers cannot contribute to the emission because 
their diffusion time scales are larger than the expansion time. 
Therefore, it is not possible to distinguish global density asymmetries
and $^{56}$Ni asymmetries based  on pre-maximum spectra.

Nevertheless, we regard off-center explosion an unlikely source 
of the polarization observed in SN~2005ke. If an off-center explosion 
is causing the early continuum polarization, the polarization should 
increase until the photosphere has receded to the Ni layers at about 
2 weeks after the explosion (i.e. around maximum light) and, then, the 
polarization angle should flip.  Our late time spectropolarimetry of SN~2005ke
does not show this effect, lending support to the conclusion that global 
density and not energy input asymmetries are at the origin of the 
continuum polarization displayed by this object.

Chemical asymmetries would cause large polarization in lines and no
continuum polarization. Therefore, in the case of subluminous events,
where the lines form far from chemical boundaries, the observed
continuum polarization requires the density distribution to be
asymmetric (see Sects.~\ref{sec:lines}, \ref{sec:continuum}).

\subsection{\label{sec:sub}Sub-Chandrasekhar Mass Explosions and Mergers}

It is still under debate whether core-normal and subluminous Type
Ia are a single class (H\"oflich et al. \cite{h02}), or form 
separate groups (Hillebrandt et al. \cite{hillebrandt10}, Sim et
al. \cite{sim11}, R\"opke et al. \cite{roepke11}).  Both
sub-Chandrasekhar mass WDs and WD mergers have been suggested 
as possible mechanisms.
 
In sub-Chandrasekhar mass models the explosion is triggered by a
detonation in the He-layers on top of a low mass WD which triggers the
detonation of the C-O core.  Explosive He burning is much more
energetic, and the resulting structures of all models show either a Ni
layer (H\"oflich \& Khokhlov \cite{hk96}; Nomoto et
al. \cite{nomoto97}; H\"oflich \cite{h97}; Woosley \cite{woosley97};
Woosley \& Kasen \cite{woosley11}) or, with some fine-tuning, strong
Fe- and Ti-rich layers (Sim et al. \cite{sim11}, Woosley et
al. \cite{woosley11}). As a generic feature, this class of models show
a $^{56}$Ni core, surrounded by a layer enriched in Si-S, subsequent
layers of O-Mg-Ne, C-O, Ni-Fe-Ti, and an outer layer of high velocity
He.  Woosley \& Kasen (\cite{woosley11}) show that maximum light
spectra look reasonable because the spectra are formed in Si/S layers
at a depth of about 0.1 $M_\odot$. Successful models must, however,
agree at all phases. In SN~2005ke at day $-$8, the spectra are formed
in the very outer layers, but we do not detect any of the chemical
signatures predicted by any of the sub-Chandrasekhar mass models.

Merger of two WDs (Iben \& Tutukov \cite{iben84}, Webbink
\cite{webbink84}; Benz et al. \cite{benz90}; Pakmor et
al. \cite{pakmor10, pakmor11}) has recently drawn attention because of
perceived advantages for solving the Type Ia progenitor problem
(Ruiter et al. \cite{ruiter09}; Schaefer \& Pagnotta
al. \cite{schaefer11}). There is evidence that mergers may form a
subclass among SNe Ia (H\"oflich \& Khokhlov \cite{hk96}; Quimby,
H\"oflich \& Wheeler \cite{quimby07}; Pakmor et al. \cite{pakmor10,
  pakmor11}).

Merger models can show a layered chemical structure similar to SD
models of $M_{Ch}$ explosions (details depend on whether or not
material from the accreted WD is assumed to detonate). There are two
aspects that qualitatively distinguish merger models from SD models:
1. late-time IR spectra and the remnant to S-Andromeda show flat
topped profiles that seem to require central densities beyond 10$^9$ g
cm$^{-3}$ in order to have electron capture time-scales comparable to
hydrodynamical timescales.  This condition is required to have a
$^{56}$Ni depletion in the center (H\"oflich et al. \cite{h04}, Maeda
et al. \cite{maeda11}); 2. merger models tend to produce a significant
amount of unburned carbon in the outer layers. Core-normal SNIa show
little or no unburned carbon (Marion et al. \cite{marion06}; Parrent
et al. \cite{parrent11}; Silverman \& Filippenko \cite{silverman12}).
These arguments argue against a merger origin for the majority of
core-normal (and SN1986g-like in the case of S And) SNe Ia. 
For the subluminous SN~2005ke, these discriminators against merger models
cannot be applied. Even explosions of single-degenerate WDs such as
DD-models for SN~1999by show strong carbon lines in the NIR down to
about 14,000 km s$^{-1}$. The existence of carbon does not sway the
argument one way or the other. Moreover, we have no late-time NIR
spectra for SN~2005ke or for any other subluminous SNe Ia and
therefore no constraint on the central density in the progenitor WDs.
Although we need a WD with a mass close to M$_{Ch}$ for SNe 1991bg,
1999by and 2005ke to explain their light curves and spectral evolution
(H\"oflich \cite{h02}), we cannot rule out an object with similar but
slightly lower mass (Mazzali et al. \cite{mazzali97},
Taubenberger et al. \cite{tauben08}). In the context of the current
work, the relatively high degree of continuum polarization observed in
the subluminous SNe 1999by and 2005ke could imply a high degree of
rotation in the progenitor and hence may be consistent with a
double-degenerate merger, for which the explosion would be
intrinsically asymmetric (e.g. Pakmor et al. \cite{pakmor11}).

We may expect that mergers induce strong differential rotation and,
thus, larger asymmetries well after maximum light (Eriguchi \&
M\"uller \cite{eriguchi85}). However, hydrodynamical models for
mergers evolved to the phase of free expansion are still missing.  In
general, we would expect large asphericity in the central region for
mergers, but almost spherical cores for rotating WDs. In light of
these considerations and the available data, we cannot rule out a
merger for SN~2005ke.

Time sequences of spectropolarimetric data and late-time NIR
spectroscopy of subluminous events may provide a more definitive
conclusion.

\section{\label{sec:concl}Conclusions}

We presented and discussed the pre-maximum
spectropolarimetry of the subluminous Type Ia event SN~2005ke. The
main observational results can be summarized as follows:

\begin{enumerate}
\item SN~2005ke shows similar properties to the only other subluminous
  event for which spectropolarimetry is available (SN~1999by). They
  differ substantially from core-normal SNe Ia.
\item Along the dominant axis the two subluminous events display a
  significant continuum polarization (0.6-0.7\% at peak) that
  increases steadily toward the red, at variance with core-normal
  events, which show a low continuum polarization ($\lesssim$0.2 \%) that is
  essentially constant in the optical band.
\item The relatively strong polarization associated with absorption
  lines in core-normal events is absent in the two subluminous objects.
\end{enumerate}

We interpreted these findings using our hydrodynamical radiation
transfer modeling. The following conclusions emerged from this
analysis:

\begin{enumerate}
\item The flux and polarization spectra of SN 2005ke are well
  reproduced by an oblate ellipsoidal geometry within a subluminous
  delayed-detonation scenario.
\item The differences with respect to core-normal SNe Ia are caused by
  low photospheric temperatures in combination with layers of unburned
  C and more massive of the products of explosive C and
  O burning.
\item The comparatively large continuum polarization is explained in
  terms of a global asymmetry ($\sim$15\%), which is not present in
  core-normal explosions
\item In the two subluminous events, the lines of intermediate mass
  elements form far from chemical boundaries and over a large velocity
  range compared to core-normal SNe Ia. This causes a blocking of the
  entire photosphere, resulting in weak line polarization.
\item The overall asphericity characterizing subluminous Type Ia may
  be produced either by a fast WD rotation, or by a double-degenerate
  merger.
\end{enumerate}

\begin{acknowledgements}
  This paper is based on observations made with ESO Telescopes at the
  Paranal Observatory under program IDs 076.D-0177(A) and
  076.D-0178(A).  The authors are grateful to ESO-Paranal staff for
  the support given during the service mode observations of
  SN~2005ke. This work is partially based on NSF grants AST
  04-06740,07-03902 \& 07-08855 to P.~A.~H., and AST-11-9801 to
  J.~C.~W., and AST-0708873 to L.~W..  The authors wish to thank an
  anonymous referee for constructive comments, which helped a lot to
  increase the clarity of the paper.
\end{acknowledgements}

\end{document}